\documentclass[journal]{IEEEtran}
\usepackage{amsmath,amsfonts}
\usepackage{algorithmic}
\usepackage{algorithm}
\usepackage{array}
\usepackage[caption=false,font=normalsize,labelfont=sf,textfont=sf]{subfig}
\usepackage{textcomp}
\usepackage{booktabs}
\usepackage{stfloats}
\usepackage{url}
\usepackage{verbatim}
\usepackage{graphicx}
\usepackage{cite}
\usepackage{soul} 
\usepackage{color, xcolor} 
\usepackage[colorlinks,
            linkcolor=cyan,
            anchorcolor=blue,
            citecolor=blue
            ]{hyperref}
\usepackage{stfloats}
\usepackage{CJKutf8}
\begin{document}
\begin{CJK}{UTF8}{gbsn}

\title{\huge From Electrode to Global Brain: Integrating \\ Multi- and Cross-Scale Brain Connections and Interactions Under Cross-Subject and Within-Subject Scenarios}
\author{Zhige Chen, and Chengxuan Qin}
\maketitle




\section{Introduction}
\label{Introduction}

Spurred on by the advent of advanced non-invasive techniques such as electroencephalogram (EEG), explorations of brain networks have entered a new era \cite{Zhao2019GraphModelingBrainNetworks}. The human brain, as a remarkable organ, exhibits a high level of time-varying complexity attributed to the intricate nature of the structural connections among its constituent units \cite{Caputi2021TopologicalAnalysis}. Such dynamic topological structure is the origin of human intelligence, contributing to understanding brain cognition activity such as motor imagery (MI) \cite{Ji2021BrainConnectivityPatterns, Saggar2022PrecisionDynamicalMapping}.\par

As determined by the interaction mechanism of the human brain \cite{Solomon2017ThetaEnhancedCognition}, the function and cognition of the whole brain can be described as a global combination of local brain regions with different topological connections and duration \cite{Sauerbrei2020CorticalPatternGeneration, Bhaduri2021AtlasofCorticalArealization, Engel2016CorticalStateSelectiveModulation}. Neurophysiological functional connectomes enable the differentiation of individuals \cite{Da2021IndividualDifferentation} and will result in the multi-scale spatial distribution difference of MI data in practical cross-subject MI classification experiments \cite{Amunts20203DProbabilisticBrainAtlas, Seeber2019SubcorticalElectrophysiologicalActivity, Genon2018CharacterizeBrainRegionFunction}, eventually leading to the global data distribution difference among subjects and reducing the classification accuracy of the classification model \cite{Mahmood2019UniversalBrainMachineInterfaces}.\par

The deep domain adaptation (DDA) method combines the superiority of deep learning and transfer learning, becoming one of the most efficient tools to address the data distribution difference problem in cross-subject EEG classification tasks \cite{Raghu2020MultiClassSeizureClassification, Cui2022EpolepsyClassification, Sun2022SubjectTransferNetwork}. More and more researchers utilize this powerful tool to solve cross-subject motor imagery (MI) classification problems \cite{Zhang2021AdaptiveTransferLearningMI, Wu2022TransferLearningMI, Wang2023MultisubjectDynamicTransfer}, aiming to improve the model generalization and the classification performance by transferring knowledge from source domain subject. \par

Despite the aforementioned advantages of deep domain adaptation (DDA) methods, the individual variabilities of EEG signals still pose great challenges to cross-subject MI classification \cite{Pernet2020IssuesRecommendationsEEG}, especially for the data-scarce single-source to single-target (STS) scenario. The existing three types of cross-subject MI classification (MTM: multi-source to multi-target, MTS: multi-source to single-target, and STS) DDA methods focus more on the global \cite{Hong2021DynamicJointAdaptationNetwork, Zhao2020JointDistributionMatching, Zhao2020RepresentationDomainAdaptation}, class \cite{Jia2023EXFineTuning, Hang2019DDAN}, and temporal domain adaptations \cite{Azab2019WeightedTransferLearning, Chen2022STSMIClassification}. While only one study explores the interplay of the electrodes in spatial domains \cite{Perez2022EEGSym}, none of them conducts the interaction of brain regions, not to mention the multi-scale spatial domain adaptation containing both electrode and region domains. \par 

In brief, no literature investigates the multi-scale spatial data distribution problem in STS cross-subject MI classification task, neither intra-subject nor inter-subject scenarios. According to the study of brain connectomics \cite{Reimann2019Micro-Connectome} and the aforementioned statement above, the topological connection of the human brain takes place on three separate levels with different scales, inextricably linked with the geometry of the brain \cite{Pang2023GeometricBrainFunction}. The proposed multi-scale spatial data distribution differences can thus be concluded as three categories under different brain scales: \par

\vspace{0.1cm} 
\begin{itemize}
    \item \textbf{\textit{(Electrode-scale)}} \textbf{1) the intra-region data distribution difference:} the data variabilities across electrodes of each brain region; \par
    \item \textbf{\textit{(Region-scale)}} \textbf{2) the inter-region data distribution difference:} consists of the data distribution pattern across brain regions under both intra-subject and inter-subject scenarios. \par
    \item \textbf{\textit{(Hemisphere-scale)}} \textbf{3) the inter-hemisphere data distribution difference:} the data distribution difference between left and right hemispheres. \par
\end{itemize}
\vspace{0.1cm} 

The multi-scale spatial data distribution differences can not be fully eliminated in motor imagery (MI) experiments for the topological structure and connection are the inherent properties of the human brain. To the best of the authors' knowledge, no previous work has integrated the multi-scale spatial data distribution problem with the deep domain adaptation network (DDAN), neither on the design of the CNN structure nor the establishment of the adaptation domain. In this paper, to integrate the principles of multi-scale brain topological structures in order to solve the multi-scale spatial data distribution difference problem \cite{Reimann2019Micro-Connectome}, a novel multi-scale spatial domain adaptation network (MSSDAN) consists of both multi-scale spatial feature extractor (MSSFE) and deep domain adaptation method called multi-scale spatial domain adaptation (MSSDA) is proposed and verified. \par

\begin{figure*}[ht]
\centering
\includegraphics[width=1\linewidth]{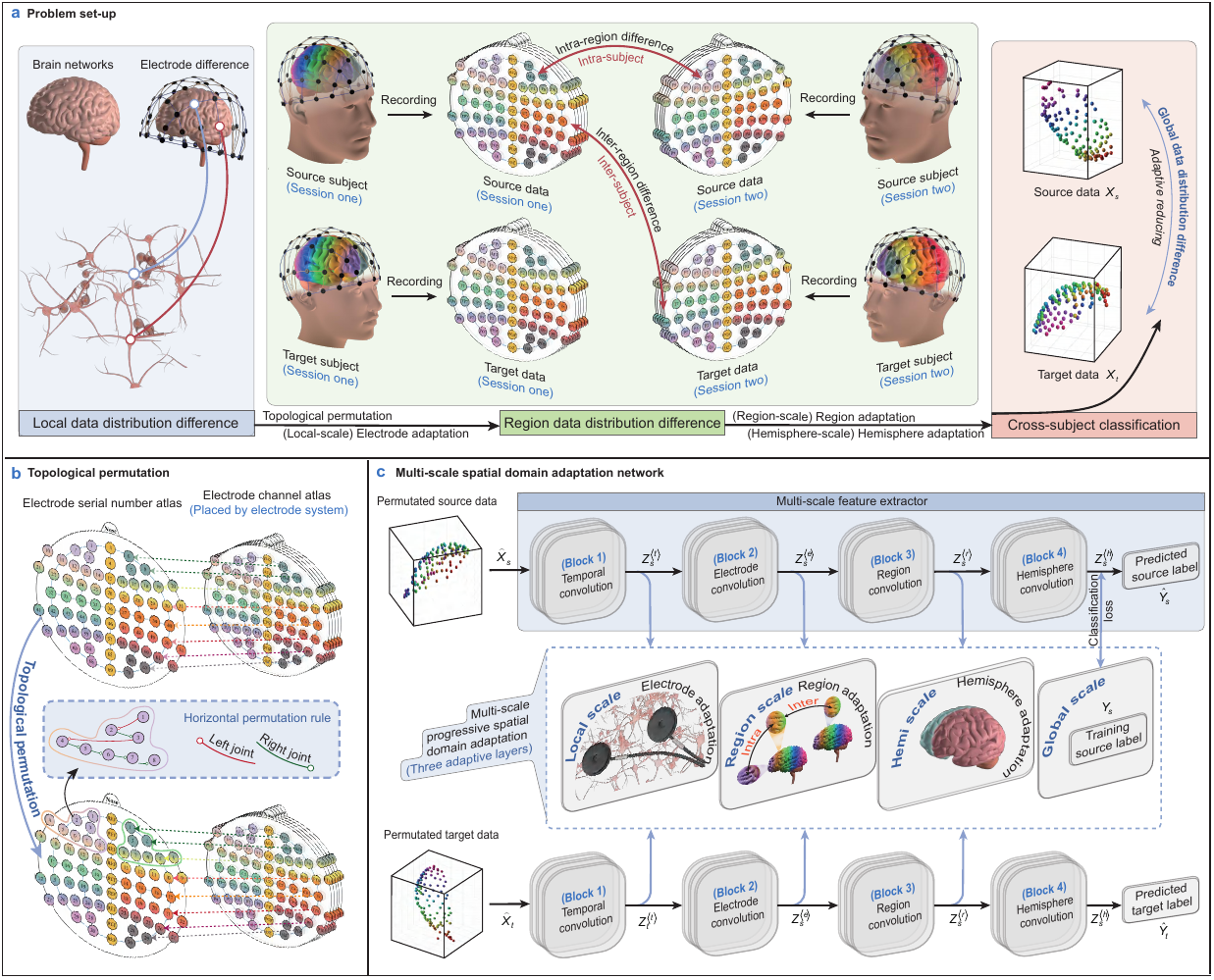}
\caption{Overview of the MSSDAN. \textbf{a}, Set-up and the flowchart of MSSDAN in STS cross-subject MI classification experiment, where multi-scale spatial data distribution differences are adaptively minimized and the unsupervised classification of the target set is accomplished. \textbf{b}, Topological permutation of the raw MI data. \textbf{c}, Architecture of the proposed domain adaptation network, MSSDAN, which consists of two key components: (1) MSSFE, a multi-scale feature extractor which extracts the deep spatial features of MI data; (2) MSSDA, multi-scale spatial domain adaptation including electrode, region, and hemisphere adaptations, transferring the spatiotemporal knowledge from the source set to the target set.}
\label{Whole_framework}
\end{figure*}

\section{Results}
\label{Results}

\subsection{Overview of MSSDAN}
In this paper, we propose MSSDAN, a new domain adaptation method for the brain-computer interface, which consists of three main components: (1) Set-up of the proposed multi-scale spatial data distribution difference problem, (2) Topological permutation and multi-scale spatial feature extractor (MSSFE), and (3) Multi-scale spatial domain adaptation (MSSDA). The whole framework in this paper is presented in Fig. \ref{Whole_framework}: our goal is to transfer the source subject's spatiotemporal knowledge (with a particular emphasis on spatial information) to the target subject to finalize the classification for the target domain. The proposed MSSDAN method is presented in the following four different aspects.\par

\subsubsection{Problem set-up}
To better understand the principles of the proposed MSSDAN model, we first describe the set-up of the proposed multi-scale spatial data distribution difference problem. As shown in the blue box of Fig. \ref{Whole_framework}a, because of the multi-scale topological structures and connections of the brain networks, the data distribution varies from electrode to electrode under both intra- and inter-subject. The signal of the electrode is the basic element of EEG signal, and as determined by the interaction mechanism of the human brain, such electrode-scale data distribution difference will thus give rise to the global data distribution difference, including region- and hemisphere-scale as shown in the green box in Fig. \ref{Whole_framework}a. \par

\begin{figure*}[ht]
\centering
\includegraphics[width=1\linewidth]{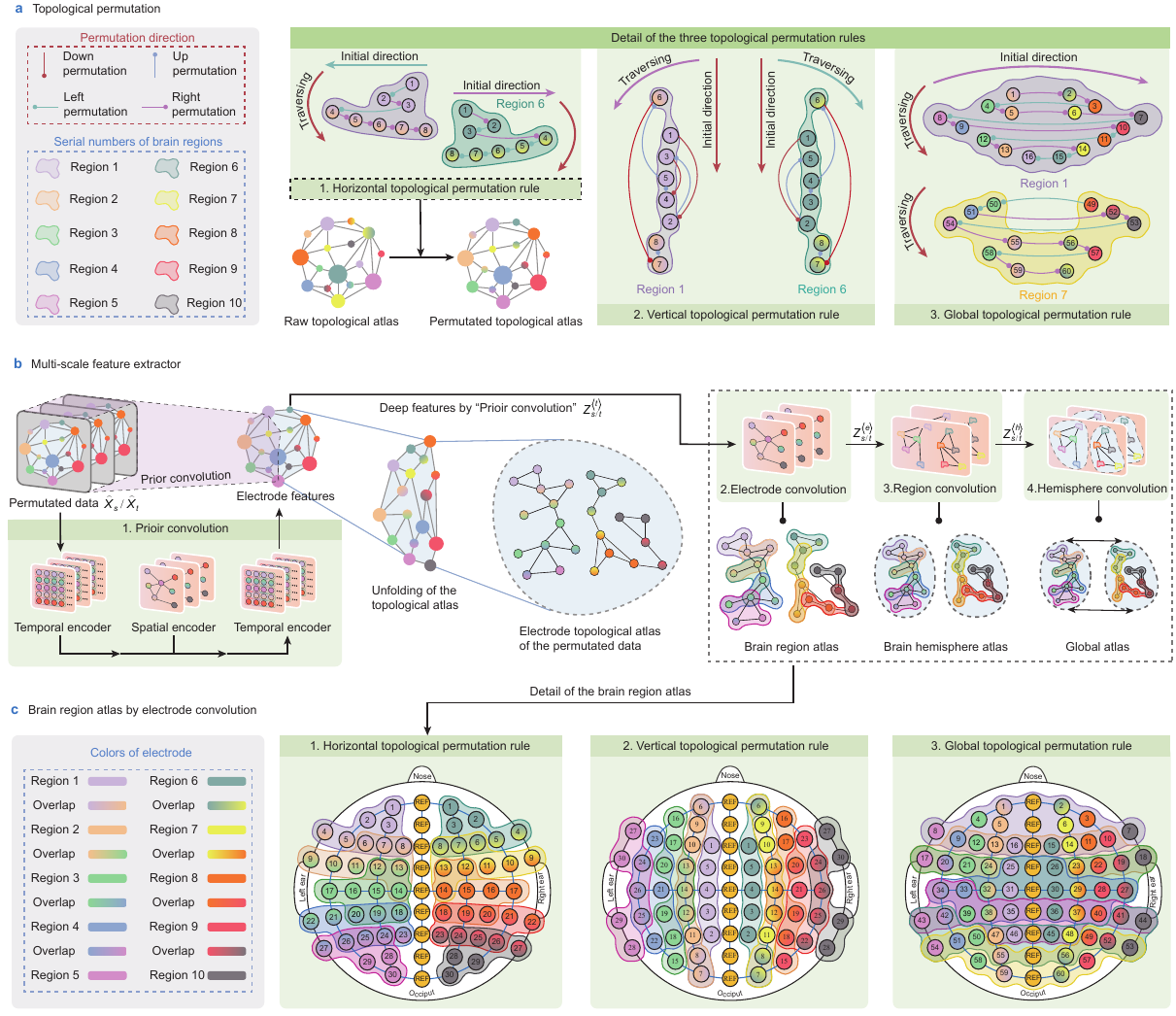}
\caption{Topological permutation and multi-scale spatial feature extractor. \textbf{a}, Topological permutation, where the electrode serials are permutated with different topological rules to facilitate the feature extraction of deep features from different brain regions. \textbf{b}, Multi-scale spatial feature extractor, where the deep features from different brain scales are extracted with customized convolution kernels and sizes. \textbf{c}, Brain region atlas obtained by electrode convolution, representing the topological properties of the permutated EEG data.}
\label{MSSFE}
\end{figure*}

As mentioned in Section \ref{Introduction}, the topological connection of the human brain can be described in three different scales, eventually leading to the global data distribution difference of inter-subject (the green box in Fig. \ref{Whole_framework}a). In order to reduce the global data distribution difference between subjects, we synthesize and utilize the data distribution differences across various brain scales, accomplishing domain adaptation across multiple brain scales (Fig. \ref{Whole_framework}c). \par

The global data distribution difference exists under both intra- and inter-subject scenarios, including the data distribution differences in both region and hemisphere scales. To solve the multi-scale spatial data distribution problem in both within- and cross-subject scenarios, the proposed domain adaptation model, MSSDAN (consists of topological permutation, MSSFE, and MSSDA as shown in Fig. \ref{Whole_framework}b and \ref{Whole_framework}c), is designed to extract the deep spatial features via a specially designed multi-scale spatial feature extractor (MSSFE), and the spatial information is transferred via multi-scale spatial domain adaptation (MSSDA). \par

As shown in the red box in Fig. \ref{Whole_framework}a, the proposed model regards two subjects' MI data sets as inputs, completing unsupervised MI classification of target domain data using source domain data and labels. The MI sets can be described as the source set $X_s=\{s(i, e)\}_{(i=1, \ldots, n_1), (e=1, \ldots, k)}$ and the target set $X_t=\{t(i, e)\}_{(j=1, \ldots, n_2), (e=1, \ldots, k)}$ with $n_1$ and $n_2$ trials respectively, where each trial consists of the electrode data from $k$ electrodes, recording the brain activity at the corresponding location. \par

Additionally, the topological brain connections are inextricably linked with the geometry of the brain, and the geographical location of the electrodes and regions are incorporated as part of the calculations. The coordinates of all the electrode channels from the source domain can be denoted as $ch_s=\left\{c_{x(e)}, c_{y(e)}\right\}_{e=1, \ldots,k}$ that contains $k$ electrodes, containing the location of the electrode in the experiment. \par

\subsubsection{Multi-scale spatial feature extractor}
With the input MI data, we first employ topological permutation to keep the raw MI data following an empirical topological permutation (Fig. \ref{MSSFE}a: a permutation that mimics the interaction of brain regions), facilitating the feature extractor to obtain via the deep features from different brain regions. \par

In this paper, we propose three typical topological permutation rules: horizontal topological permutation rule, vertical topological permutation rule, and global topological permutation rule. All the topological rules are designed and computed in terms of the spatial properties of the EEG signals, assigning topological information to the raw electrode series. The permuted MI data from both two sets are then loaded into the multi-scale feature extractor, obtaining the deep spatial features in different brain scales (electrode-scale, region-scale, and hemisphere-scale) as shown in Fig. \ref{MSSFE}b. \par

As shown in Fig. \ref{MSSFE}b, the specially designed multi-scale spatial feature extractor consists of four convolution blocks, extracting spatial features from different brain scales. With the empirical topological information assigned by different topological permutation rules, the electrode will be divided into different brain regions, with different brain region atlas shown in Fig. \ref{MSSFE}c.

\subsubsection{Multi-scale spatial domain adaptation}
With the multi-scale spatial features extracted by MSSFE, the spatiotemporal knowledge from the source set is then transferred to the target set via MSSDA. The MSSDA incorporates three adaptation layers with different spatial scales: electrode adaptation, region adaptation, and hemisphere adaptation, integrating multi-scale and cross-scale brain connections and interactions under cross-subject and within-subject scenarios. \par

\section{Code Availability}
The source code of this work is available after peer review.

\end{CJK}

\begin{thebibliography}{00}
\bibitem{Amunts20203DProbabilisticBrainAtlas}
K.~Amunts, H.~Mohlberg, S.~Bludau, and K.~Zilles, ``Julich-brain: A 3{D} probabilistic atlas of the human brain’s cytoarchitecture, ''\emph{Science}, vol. 369, no. 6506, pp. 988--992, 2020.

\bibitem{Azab2019WeightedTransferLearning}
A.~M. Azab, L.~Mihaylova, K.~K. Ang, and M.~Arvaneh, ``Weighted transfer learning for improving motor imagery-based brain-computer interface, ''\emph{IEEE Transactions on Neural Systems and Rehabilitation Engineering}, vol.~27, no.~7, pp. 1352--1359, 2019.

\bibitem{Bhaduri2021AtlasofCorticalArealization}
A.~Bhaduri, C.~Sandoval-Espinosa, M.~Otero-Garcia, I.~Oh, R.~Yin, U.~C. Eze, T.~J. Nowakowski, and A.~R. Kriegstein, ``An atlas of cortical arealization identifies dynamic molecular signatures,'' \emph{Nature}, vol. 598, no. 7879, pp. 200--204, 2021.

\bibitem{Caputi2021TopologicalAnalysis}
L.~Caputi, A.~Pidnebesna, and J.~Hlinka, ``Promises and pitfalls of topological data analysis for brain connectivity analysis,'' \emph{NeuroImage}, vol. 238, p. 118245, 2021.

\bibitem{Chen2022STSMIClassification}
Y.~Chen, R.~Yang, M.~Huang, Z.~Wang, and X.~Liu, ``Single-source to single-target cross-subject motor imagery classification based on multisubdomain adaptation network,'' \emph{IEEE Transactions on Neural Systems and Rehabilitation Engineering}, vol.~30, pp. 1992--2002, 2022.

\bibitem{Chen2024EEGprogress}
Z.~Chen, R.~Yang, M.~Huang, F.~Li, G.~Lu, and Z.~Wang, ``{EEGP}rogress: {A} fast and lightweight progressive convolution architecture for {EEG} classification,'' \emph{Computers in Biology and Medicine}, vol. 169, p. 107901, 2024.
  
\bibitem{Chen2023EDAN}
Z.~Chen, R.~Yang, M.~Huang, Z.~Wang, and X.~Liu, ``Electrode domain adaptation network: {M}inimizing the difference across electrodes in single-source to single-target motor imagery classification,'' \emph{IEEE Transactions on Emerging Topics in Computational Intelligence}, 2023.

\bibitem{chen2024bdan}
Z.~Chen, R.~Yang, M.~Huang, C.~Qin, and Z.~Wang, ``B{DAN}: Mitigating temporal difference across electrodes in cross-subject motor imagery classification via generative bridging domain,'' \emph{arXiv preprint arXiv:2404.10494}, 2024.




\bibitem{Cui2022EpolepsyClassification}
X.~Cui, D.~Hu, P.~Lin, J.~Cao, X.~Lai, T.~Wang, T.~Jiang, and F.~Gao, ``Deep feature fusion based childhood epilepsy syndrome classification from electroencephalogram,'' \emph{Neural Networks}, vol. 150, pp. 313--325, 2022.

\bibitem{Da2021IndividualDifferentation}
J.~da~Silva~Castanheira, H.~D. Orozco~Perez, B.~Misic, and S.~Baillet, ``Brief segments of neurophysiological activity enable individual differentiation,'' \emph{Nature Communications}, vol.~12, no.~1, p. 5713, 2021.

\bibitem{Engel2016CorticalStateSelectiveModulation}
T.~A. Engel, N.~A. Steinmetz, M.~A. Gieselmann, A.~Thiele, T.~Moore, and K.~Boahen, ``Selective modulation of cortical state during spatial attention,'' \emph{Science}, vol. 354, no. 6316, pp. 1140--1144, 2016.

\bibitem{Fang2023FastRecon}
Z.~Fang, X.~Wang, H.~Li, J.~Liu, Q.~Hu, and J.~Xiao, ``Fastrecon: {F}ew-shot industrial anomaly detection via fast feature reconstruction,'' in \emph{Proceedings of the IEEE/CVF International Conference on Computer Vision}, 2023, pp.~17481--17490.

\bibitem{Genon2018CharacterizeBrainRegionFunction}
S.~Genon, A.~Reid, R.~Langner, K.~Amunts, and S.~B. Eickhoff, ``How to characterize the function of a brain region,'' \emph{Trends in Cognitive Sciences}, vol.~22, no.~4, pp. 350--364, 2018.

\bibitem{Hang2019DDAN}
W.~Hang, W.~Feng, R.~Du, S.~Liang, Y.~Chen, Q.~Wang, and X.~Liu, ``Cross-subject {EEG} signal recognition using deep domain adaptation network,'' \emph{IEEE Access}, vol.~7, pp. 128\,273--128\,282, 2019.

\bibitem{Hong2021DynamicJointAdaptationNetwork}
X.~Hong, Q.~Zheng, L.~Liu, P.~Chen, K.~Ma, Z.~Gao, and Y.~Zheng, ``Dynamic joint domain adaptation network for motor imagery classification,'' \emph{IEEE Transactions on Neural Systems and Rehabilitation Engineering},  vol.~29, pp. 556--565, 2021.

\bibitem{Huang2020MultiTaskLearning}
Z.-A.~Huang, R.~Liu, and K.~C.~Tan, ``Multi-task learning for efficient diagnosis of {ASD} and {ADHD} using resting-state f{MRI} data,'' in \emph{2020 International Joint Conference on Neural Networks (IJCNN)}, 2020, pp.~1--7.

\bibitem{Jiang2021Optimization}
Z.~Jiang, Y.~Guo, K.~Jiang, M.~Hu, and Z.~Zhu, ``Optimization of intelligent plant cultivation robot system in object detection,'' \emph{IEEE Sensors Journal}, vol.~21, no.~17, pp.~19279--19288, 2021.

\bibitem{Jiang2024SocialNSTransformers}
Z.~Jiang, Y.~Ma, B.~Shi, X.~Lu, J.~Xing, N.~Gon{\c{c}}alves, and B.~Jin, ``Social {NST}ransformers: {L}ow-Quality Pedestrian Trajectory Prediction,'' \emph{IEEE Transactions on Artificial Intelligence}, 2024.

\bibitem{Ji2021BrainConnectivityPatterns}
J.~Ji, X.~Xing, Y.~Yao, J.~Li, and X.~Zhang, ``Convolutional kernels with an element-wise weighting mechanism for identifying abnormal brain connectivity patterns,'' \emph{Pattern Recognition}, vol. 109, p. 107570, 2021.

\bibitem{Jia2023EXFineTuning}
X.~Jia, Y.~Song, and L.~Xie, ``Excellent fine-tuning: {F}rom specific-subject classification to cross-task classification for motor imagery,'' \emph{Biomedical Signal Processing and Control}, vol.~79, p. 104051, 2023.

\bibitem{Liu2022AttentionLikeFusion}
R.~Liu, Z.-A.~Huang, Y.~Hu, Z.~Zhu, K.-C.~Wong, and K.~C.~Tan, ``Attention-like multimodality fusion with data augmentation for diagnosis of mental disorders using {MRI},'' \emph{IEEE Transactions on Neural Networks and Learning Systems}, 2022.

\bibitem{Mahmood2019UniversalBrainMachineInterfaces}
M.~Mahmood, D.~Mzurikwao, Y.-S. Kim, Y.~Lee, S.~Mishra, R.~Herbert, A.~Duarte, C.~S. Ang, and W.-H. Yeo, ``Fully portable and wireless universal brain--machine interfaces enabled by flexible scalp electronics and deep learning algorithm,'' \emph{Nature Machine Intelligence}, vol.~1, no.~9, pp. 412--422, 2019.

\bibitem{Pang2023GeometricBrainFunction}
J.~C. Pang, K.~M. Aquino, M.~Oldehinkel, P.~A. Robinson, B.~D. Fulcher, M.~Breakspear, and A.~Fornito, ``Geometric constraints on human brain function,'' \emph{Nature}, pp. 1--9, 2023.

\bibitem{Perez2022EEGSym}
S.~P{\'e}rez-Velasco, E.~Santamar{\'\i}a-V{\'a}zquez, V.~Mart{\'\i}nez-Cagigal, D.~Marcos-Mart{\'\i}nez, and R.~Hornero, ``{EEGSym}: {O}vercoming inter-subject variability in motor imagery based {BCI}s with deep learning,'' \emph{IEEE Transactions on Neural Systems and Rehabilitation Engineering}, vol.~30, pp. 1766--1775, 2022.

\bibitem{Pernet2020IssuesRecommendationsEEG}
C.~Pernet, M.~I. Garrido, A.~Gramfort, N.~Maurits, C.~M. Michel, E.~Pang, R.~Salmelin, J.~M. Schoffelen, P.~A. Valdes-Sosa, and A.~Puce, ``Issues and recommendations from the ohbm cobidas {MEEG} committee for reproducible {EEG} and {MEG} research,'' \emph{Nature Neuroscience}, vol.~23, no.~12, pp. 1473--1483, 2020.

\bibitem{Qin2024EEGUnity}
C.~Qin, R.~Yang, W.~You, Z.~Chen, L.~Zhu, M.~Huang, and Z.~Wang, ``{EEGU}nity: {O}pen-source tool in facilitating unified {EEG} datasets towards large-scale {EEG} model,'' \emph{arXiv preprint arXiv:2410.07196}, 2024.

\bibitem{Qin2023SpatialVariation}
C.~Qin, R.~Yang, M.~Huang, W.~Liu, and Z.~Wang, ``Spatial variation generation algorithm for motor imagery data augmentation: {I}ncreasing the density of sample vicinity,'' \emph{IEEE Transactions on Neural Systems and Rehabilitation Engineering}, 2023.

\bibitem{Raghu2020MultiClassSeizureClassification}
S.~Raghu, N.~Sriraam, Y.~Temel, S.~V. Rao, and P.~L. Kubben, ``{EEG} based multi-class seizure type classification using convolutional neural network and transfer learning,'' \emph{Neural Networks}, vol. 124, pp. 202--212,
 2020.

\bibitem{Reimann2019Micro-Connectome}
M.~W. Reimann, M.~Gevaert, Y.~Shi, H.~Lu, H.~Markram, and E.~Muller, ``A null model of the mouse whole-neocortex micro-connectome,'' \emph{Nature Communications}, vol.~10, no.~1, p. 3903, 2019.

\bibitem{Saggar2022PrecisionDynamicalMapping}
M.~Saggar, J.~M. Shine, R.~Li{\'e}geois, N.~U. Dosenbach, and D.~Fair, ``Precision dynamical mapping using topological data analysis reveals a hub-like transition state at rest,'' \emph{Nature Communications}, vol.~13,  no.~1, p. 4791, 2022.

\bibitem{Sauerbrei2020CorticalPatternGeneration}
B.~A. Sauerbrei, J.-Z. Guo, J.~D. Cohen, M.~Mischiati, W.~Guo, M.~Kabra, N.~Verma, B.~Mensh, K.~Branson, and A.~W. Hantman, ``Cortical pattern generation during dexterous movement is input-driven,'' \emph{Nature}, vol. 577, no. 7790, pp. 386--391, 2020.

\bibitem{Seeber2019SubcorticalElectrophysiologicalActivity}
M.~Seeber, L.-M. Cantonas, M.~Hoevels, T.~Sesia, V.~Visser-Vandewalle, and C.~M. Michel, ``Subcortical electrophysiological activity is detectable with high-density {EEG} source imaging,'' \emph{Nature Communications}, vol.~10, no.~1, p. 753, 2019.

\bibitem{Solomon2017ThetaEnhancedCognition}
E.~A. Solomon, J.~E. Kragel, M.~R. Sperling, A.~Sharan, G.~Worrell, M.~Kucewicz, C.~S. Inman, B.~Lega, K.~A. Davis, J.~M. Stein \emph{et~al.}, ``Widespread theta synchrony and high-frequency desynchronization underlies enhanced cognition,'' \emph{Nature Communications}, vol.~8, no.~1, p. 1704, 2017.

\bibitem{Sun2022SubjectTransferNetwork}
B.~Sun, Z.~Wu, Y.~Hu, and T.~Li, ``Golden subject is everyone: {A} subject transfer neural network for motor imagery-based brain computer interfaces,'' \emph{Neural Networks}, vol. 151, pp. 111--120, 2022.

\bibitem{Wang2023MultisubjectDynamicTransfer}
H.~Wang, P.~Chen, M.~Zhang, J.~Zhang, X.~Sun, M.~Li, X.~Yang, and Z.~Gao, ``E{EG}-based motor imagery recognition framework via multisubject dynamic transfer and iterative self-training,'' \emph{IEEE Transactions on Neural Networks and Learning Systems}, 2023.

\bibitem{Wang2024MMPT}
T.~Wang, Y.~Liu, J.~C.~Liang, Y.~Cui, Y.~Mao, S.~Nie, J.~Liu, F.~Feng, Z.~Xu, C.~Han, \emph{et~al.}, ``{MMPT}: {M}ultimodal prompt tuning for zero-shot instruction learning,'' \emph{arXiv preprint arXiv:2409.15657}, 2024.

\bibitem{Wu2022TransferLearningMI}
D.~Wu, X.~Jiang, and R.~Peng, ``Transfer learning for motor imagery based brain--computer interfaces: {A} tutorial,'' \emph{Neural Networks}, vol. 153, pp. 235--253, 2022.

\bibitem{Zhang2021AdaptiveTransferLearningMI}
K.~Zhang, N.~Robinson, S.-W. Lee, and C.~Guan, ``Adaptive transfer learning for EEG motor imagery classification with deep convolutional neural network,'' \emph{Neural Networks}, vol. 136, pp. 1--10, 2021.

\bibitem{Zhao2020RepresentationDomainAdaptation}
H.~Zhao, Q.~Zheng, K.~Ma, H.~Li, and Y.~Zheng, ``Deep representation-based domain adaptation for nonstationary {EEG} classification,'' \emph{IEEE Transactions on Neural Networks and Learning Systems}, vol.~32, no.~2, pp. 535--545, 2020.

\bibitem{Zhao2019GraphModelingBrainNetworks}
T.~Zhao, Y.~Xu, and Y.~He, ``Graph theoretical modeling of baby brain networks,'' \emph{NeuroImage}, vol. 185, pp. 711--727, 2019.

\bibitem{Zhao2020JointDistributionMatching}
X.~Zhao, J.~Zhao, C.~Liu, W.~Cai \emph{et~al.}, ``Deep neural network with joint distribution matching for cross-subject motor imagery brain-computer interfaces,'' \emph{BioMed Research International}, vol. 2020, 2020.

\end{thebibliography}
\end{document}